\documentclass[copyright,creativecommons]{eptcs}

\providecommand{\event}{ACL2-2020}

\usepackage{alltt}
\usepackage{amssymb}
\usepackage{enumitem}
\usepackage[T1]{fontenc}
\usepackage{hyperref}
\usepackage{inconsolata}
\usepackage{MnSymbol}
\usepackage{tikz}
\usepackage{underscore}
\usepackage{url}
\usepackage{xcolor}
\usepackage[all]{xy}

\newcommand{\spec}{s}
\newcommand{\refsto}{\leadsto}
\newcommand{\refsfrom}{\leftlsquigarrow}
\newcommand{\tocode}
 {\stackrel{\mbox{\raisebox{-1pt}{\scriptsize$\mathbf{c}$}}}{\mapsto}}
\newcommand{\fromcode}
 {\stackrel{\mbox{\raisebox{-1pt}{\scriptsize$\mathbf{r}$}}}{\mapsto}}
\newcommand{\prog}{p}

\newcommand{\funfromto}[3]{#1:#2\rightarrow#3}
\newcommand{\funinv}[1]{#1^{-1}}
\newcommand{\funcomp}[2]{#1\circ#2}
\newcommand{\funcompcomp}[3]{#1\circ#2\circ#3}
\newcommand{\funid}[1]{\mathit{id}_{#1}}
\newcommand{\fundef}{\equiv}
\newcommand{\ifte}[3]{\ \mathbf{if}\ #1\ \mathbf{then}\ #2\ \mathbf{else}\ #3}
\newcommand{\meas}{\mu}
\newcommand{\wfrel}{\prec}
\newcommand{\isoin}{\xi}
\newcommand{\osiin}{\funinv{\isoin}}
\newcommand{\isoout}{\upsilon}
\newcommand{\osiout}{\funinv{\isoout}}

\newcommand{\acl}[1]{\texttt{#1}} 
\newenvironment{bacl}{\small\begin{alltt}}{\end{alltt}} 

\newcommand{\secref}[1]{Section \ref{#1}}
\newcommand{\figref}[1]{Figure \ref{#1}}

\newcommand{\citeman}[2]{\cite[Topic \href{#1}{\texttt{#2}}]{acl2-manual}}

\begin{document}


\title{Isomorphic Data Type Transformations}

\author{Alessandro Coglio
        \institute{Kestrel Institute \\ \url{http://www.kestrel.edu}}
        \and
        Stephen Westfold
        \institute{Kestrel Institute \\ \url{http://www.kestrel.edu}}}

\def\titlerunning{Isomorphic Data Type Transformations}
\def\authorrunning{A. Coglio and S. Westfold}

\maketitle

\begin{abstract}

In stepwise derivations of programs from specifications, data type refinements
are common. Many data type refinements
involve isomorphic mappings between the more abstract
and more concrete data representations.  Examples include refinement of finite
sets to duplicate-free ordered lists or to bit vectors, adding record
components that are functions of the other fields to avoid expensive
recomputation, etc.
This paper describes the APT (Automated Program Transformations)
tools to carry out isomorphic data type refinements in the ACL2 theorem prover
and gives examples of their use.  Because of the inherent symmetry of
isomorphisms, these tools are also useful to verify existing programs, by
turning more concrete data representations into more abstract ones to ease
verification.  Typically, a data type will have relatively few interface
functions that access the internals of the type.  Once versions of these
interface functions have been derived that work on the isomorphic type,
higher-level functions can be derived simply by substituting the old functions
for the new ones.  We have implemented the APT transformations \acl{isodata} to
generate the former, and \acl{propagate-iso} for generating the latter functions
as well as theorems about the generated functions from the theorems about the
original functions.  \acl{Propagate-iso} also handles cases where the type is a
component of a more complex one such as a list of the type or a record that has
a field of the type: the isomorphism on the component type is automatically
lifted to an isomorphism on the more complex type.  As with all APT
transformations, \acl{isodata} and \acl{propagate-iso} generate proofs of the
relationship of the transformed functions to the originals.

\end{abstract}



\section{Background, Motivation, and Contribution}
\label{sec:intro}

In stepwise program refinement~\cite{dijkstra-constructive,wirth-refinement},
an implementation is derived from a specification
via a sequence of intermediate specifications.
A derivation is a sequence
$\spec_0 \refsto \spec_1 \refsto \ldots \refsto \spec_n \tocode \prog$,
where:
$\spec_0$ is a requirements specification in some specification language;
$\spec_1,\ldots,\spec_n$ are intermediate specifications in the same language;
$\refsto$ is a refinement relation (a preorder, i.e.\ reflexive and transitive);
and $\prog$ is an implementation in some programming language,
obtained from $\spec_n$ via a code generator $\tocode$.
If the specification language is a superset of a programming language,
code generation may be omitted and $\prog = \spec_n$.

If every derivation step (including code generation if applicable)
is formally verified,
the implementation is provably correct by construction.
Given $\spec_i$, the step $\spec_i \refsto \spec_{i+1}$
may be realized in two ways.
One is by writing $\spec_{i+1}$ and verifying the $\refsto$ relation;
this is `posit-and-prove' \cite{bmethod,zed,vdm}.
The other (when possible) is
by automatically generating both $\spec_{i+1}$ and the $\refsto$ proof
from $\spec_i$ via an \emph{automated transformation}
\cite{kids,smith-marktoberdorf,specware-www,simplify,apt-www}.
The latter approach may require the developer
to prove \emph{applicability conditions},
i.e.\ theorems from which $\refsto$ can then be proved automatically;
but proving applicability conditions is generally easier than proving $\refsto$.
Code generation $\spec_n\tocode\prog$ is normally automatic;
its verification is akin to compilation verification.

Each step $\spec_i \refsto \spec_{i+1}$
narrows down the possible implementations
or rephrases the specification towards an implementation,
e.g.\ by choosing an algorithm or data structure,
or by exploiting algebraic laws to optimize expressions.
A common case is that of a data type refinement,
where a more abstract data representation
is turned into a more concrete one \cite{hoare-data}.
A large and interesting subset of data type refinements
involves isomorphic mappings between abstract and concrete representations;
examples are
refining finite sets to duplicate-free ordered lists or to bit vectors,
adding record components that are computable from the others for caching,
and some perhaps less expected ones as in \secref{sec:loop}.
Many other data type refinements do not involve isomorphic mappings;
see \secref{sec:future}.

In an interactive theorem prover like ACL2,
stepwise program refinement can be carried out
using predicates over (deeply or shallowly embedded) implementations
as specifications $\spec_i$
and (reversed) implication
as $\refsto$
(an approach called `pop-refinement') \cite{popref,soft};
and $\tocode$ may be realized
either as a $\refsto$ sequence entirely in the logic of the prover \cite{popref},
or via a more typical code generator \cite{aij-atj}
\citeman{http://www.cs.utexas.edu/users/moore/acl2/manuals/current/manual/?topic=JAVA____AT}{ATJ}.
The APT (Automated Program Transformations) library
\citeman{http://www.cs.utexas.edu/users/moore/acl2/manuals/current/manual/?topic=APT____APT}{APT}
provides tools to carry out derivation steps $\spec_i\refsto\spec_{i+1}$
via automated transformations (as outlined above) in ACL2.
This paper describes the APT tools
to carry out isomorphic data type refinements via transformations.

Besides deriving programs from specifications,
APT is useful to verify existing code.
One can build an \emph{anti-derivation},
i.e.\ a sequence
$\prog \fromcode \spec_0' \refsfrom \spec_1' \refsfrom \ldots \refsfrom\spec_m'$,
where:
$\prog$ is an existing program;
$\fromcode$ yields a representation $\spec_0'$ of the program
in the logical specification language;
$\refsfrom$ is an \emph{anti-refinement} relation (converse of $\refsto$);
and $\spec_m'$ is a more abstract representation of the program
that should be easier to verify than $\spec_0'$.
In the \emph{analysis-by-synthesis} approach
\cite{analysis-by-synthesis,derivationminer-www},
an anti-derivation
$\prog \fromcode \spec_0' \refsfrom \spec_1' \refsfrom \ldots \refsfrom\spec_m'$
can be combined with a derivation without code generation
$\spec_0 \refsto \spec_1 \refsto \ldots \refsto \spec_n$,
where $\spec_n$ and $\spec_m'$ are equal or trivially equivalent,
to verify that $\prog$ satisfies $\spec_0$.
A step $\spec_j' \refsfrom \spec_{j+1}'$ in an anti-derivation
may be carried out (when possible)
via an automated transformation that is ``inverse'' of
one that would realize the converse step $\spec_{j+1}' \refsto \spec_j'$
in a derivation.
The isomorphic data type transformations described in this paper
are readily invertible, due to the inherent symmetry of isomorphisms.
(Other kinds of transformations are more difficult to invert.)


\section{Mathematical Overview}
\label{sec:overview}

Here functions are viewed
not extensionally as possibly infinite sets of pairs,
but intensionally as finite descriptions such as
the function definitions in ACL2 and similar languages.

Consider a function $\funfromto{f}{X}{Y}$
with definition $f(x) \fundef \ifte{a(x)}{b(x)}{c(x,f(d(x)))}$,
where:
$a(x)$ is the termination test;
$b(x)$ is the base case of the recursion;
$c(x,f(d(x)))$ combines the recursive call with the argument;
and $d(x)$ decreases some measure $\meas$ of the argument
according to some well-founded relation $\wfrel$,
i.e.\ $\neg\; a(x) \implies \meas(d(x)) \wfrel \meas(x)$,
so that $f$ is logically consistent.
This is a representative recursive function definition,
which covers special cases such as non-recursion (when $a(x)$ is always true),
and whose generalization to multiple base cases and recursive calls
is easily imagined.
When $a$, $b$, $c$, and $d$ are executable,
$f$ describes a computation
from inputs in $X$ to outputs in $Y$.

Consider an isomorphism $\funfromto{\isoin}{X}{X'}$
with inverse $\funfromto{\osiin}{X'}{X}$
(so $\funcomp{\osiin}{\isoin} = \funid{X}$
and $\funcomp{\isoin}{\osiin} = \funid{X'}$),
and an isomorphism $\funfromto{\isoout}{Y}{Y'}$
with inverse $\funfromto{\osiout}{Y'}{Y}$
(so $\funcomp{\osiout}{\isoout} = \funid{Y}$
and $\funcomp{\isoout}{\osiout} = \funid{Y'}$).
If $\isoin$ and $\isoout$ are executable,
they describe computations
to change the representations of $f$'s inputs and outputs
to the ones in $X'$ and $Y'$.
If $\osiin$ and $\osiout$ are also executable,
they describe computations to change the representations
back to the ones in $X$ and $Y$.

An \emph{isomorphic} version $\funfromto{f'}{X'}{Y'}$ of $f$
that operates on the values in $X'$ and $Y'$
can be mechanically defined as
$f'(x') \fundef
 \ifte{a(\osiin(x'))}
      {\isoout(b(\osiin(x')))}
      {\isoout(c(\osiin(x'),\osiout(f'(\isoin(d(\osiin(x')))))))}$,
which has the same form as $f$
but with the four isomorphic conversions ($\isoin$, $\isoout$, etc.)
added to make it ``type-correct''.%
\footnote{This covers the special case in which
only the output data representation is changed
(i.e.\ $X' = X$ and $\isoin = \osiin = \funid{X}$),
or only the input data representation is changed
(i.e.\ $Y' = Y$ and $\isoout = \osiout = \funid{Y}$).
It also covers the case in which $f$ has
multiple inputs
(i.e.\ $X = X_1 \times \cdots \times X_n$
and $X' = X_1' \times \cdots \times X_{n'}'$,
with either $n = n'$ or $n \neq n'$)
and/or multiple outputs
(i.e.\  $Y = Y_1 \times \cdots \times Y_m$
and $Y' = Y_1' \times \cdots \times Y_{m'}'$
with either $m = m'$ or $m \neq m'$).}
The relationship between $f'$ and $f$ is $f' = \funcompcomp{\isoout}{f}{\osiin}$,
or equivalently $f = \funcompcomp{\osiout}{f'}{\isoin}$,
illustrated in the commuting diagram in \figref{fig:commuting};
this is proved by induction, using the properties of the isomorphisms.
The logical consistency of $f'$ follows from the one of $f$,
using the measure $\funcomp{\meas}{\osiin}$.
The computation described by $f'$ is the same as $f$,
with the addition of isomorphic conversions back and forth as needed.

\begin{figure}
\centering
\ 
\xymatrix@=10ex{
X \ar[r]^f \ar@<0.5ex>[d]^\isoin & Y \ar@<0.5ex>[d]^\isoout \\
X' \ar[r]^{f'} \ar@<0.5ex>[u]^\osiin & Y' \ar@<0.5ex>[u]^\osiout
}
\caption{Relationship between $f$ and its isomorphic version $f'$.%
\label{fig:commuting}}
\end{figure}
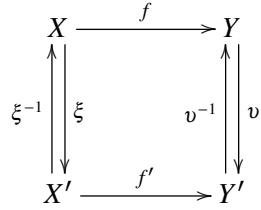

Expanding the definitions of $a$, $b$, $c$, $d$, $\isoin$, etc.,
it may be possible to use rewrite rules to simplify the body of $f'$,
obtaining an equivalent new function definition
that operates directly on $X'$ and $Y'$
and no longer uses the isomorphic conversions and the old data representations.
This is a separate process,
which may require user guidance in general.
But it can be automated
if there exist (recursively obtained%
\footnote{Here `recursively' refers to the call graph of the functions.})
isomorphic versions
$a'$ of $a$ with $a(x) = a'(\isoin(x))$,%
\footnote{Since here $a$ and $a'$ return booleans,
only the representation of their inputs is changed.}
$b'$ of $b$ with $b(x) = \osiout(b'(\isoin(x)))$,
$c'$ of $c$ with $c(x,y) = \osiout(c'(\isoin(x),\isoout(y)))$, and
$d'$ of $d$ with $d(x) = \osiin(d'(\isoin(x))$.
The latter four equalities, along with the properties of the isomorphisms,
applied as rewrite rules directly in the definition of $f'$,
yield $f'(x') \fundef \ifte{a'(x')}{b'(x')}{c'(x',f'(d'(x')))}$.
The automation of this process is enabled by the fact that
the definition of $f$ manipulates values in $X$ and $Y$
exclusively via functions ($a$ etc.)
that have isomorphic versions ($a'$ etc.).%
\footnote{This would be still the case if the definition of $f$
also manipulated values in $X$ and $Y$ with certain functions, like equality,
that operate ``polymorphically'' on all value representations,
without the need to have isomorphic versions for these functions.}
In this case,
the definition of $f'$ without the isomorphic conversions
can be mechanically constructed from the definition of $f$,
and so can the proofs of the relationship between $f$ and $f'$ as above.
We say that the isomorphic transformations of $a$, $b$, etc.\
are \emph{propagated} to $f$.
If instead not all the functions in $f$ that manipulate values in $X$ and $Y$
have isomorphic versions,
we say that we \emph{initiate} the isomorphic transformation at $f$.
If some but not all of the functions $a$, $b$, etc.\ have isomorphic versions,
their associated isomorphic relationships
can be used in the subsequent user-guided simplification of $f'$,
but this case is still considered an initiation and not a propagation.
In a well-engineered development
where the data structures whose representations are being transformed
are accessed via a relatively small number of interface functions,
isomorphic transformations will be
initiated at the interface functions
and propagated to all the other functions.

In general, a function like $f$ above may be part of
(i.e.\ directly or indirectly called by)
a requirements or intermediate specification $\spec_i$
of the kind described in \secref{sec:intro}.
For instance, a requirements or intermediate specification of a compiler
may involve functions that manipulate
data structures like stacks and symbol tables,
which may be transformed into more efficient isomorphic data structures
as part of a derivation of an implementation of the compiler.
There may be several functions ($f$, $a$, etc.)
manipulating these data structures.
In subjecting all these functions to isomorphic transformations,
some will require user-guided simplification
to eliminate the isomorphic conversions,
but others will be amenable to automatic propagation,
including the predicate $\spec_i$.%
\footnote{If $X$ and $Y$ are part of the required ``interface'' in $\spec_0$
of the program being derived,
note that a wrapper $\widetilde{f} \fundef \funcompcomp{\osiout}{f'}{\isoin}$
can be mechanically defined along with $f'$,
provably satisfying $\widetilde{f} = f$.
This is different from the definition of $f'$
that includes the isomorphic conversions:
$\widetilde{f}$ only converts the input once and the output once,
while the definition of $f'$ that includes the isomorphic conversions
performs several conversions at each recursive call.}


\section{Realization in ACL2}
\label{sec:tools}

We developed three ACL2 tools to realize
the isomorphic transformations described in \secref{sec:overview}:
\begin{enumerate}[nosep]
\item
\acl{Defiso}, to establish isomorphic mappings
like $\xymatrix@1{X\ar[r]|{\;\isoin\;} & X'}$
and $\xymatrix@1{Y\ar[r]|{\;\isoout\;} & Y'}$.
See \secref{sec:defiso}.
\item
\acl{Isodata}, to initiate isomorphic transformations
based on \acl{defiso} mappings.
See \secref{sec:isodata}.
\item
\acl{Propagate-iso}, to propagate isomorphic transformations
initiated by \acl{isodata}.
See \secref{sec:propiso}.
\end{enumerate}


\subsection{Establishing Isomorphic Mappings}
\label{sec:defiso}

Even though the ACL2 language is untyped,
predicates (i.e.\ functions that are, or are treated as, boolean-valued)
often play the role of types in ACL2.
Thus, sets of data representations like $X$, $X'$ etc.\ in \secref{sec:overview}
are denoted by ACL2 predicates.
Isomorphisms like $\isoin$, $\osiin$, etc.\ in \secref{sec:overview}
are denoted by ACL2 functions.
Since functions are total in ACL2 (i.e.\ always defined over all values),
the isomorphisms are ``relativized'' to the predicates,
as explicated below.

\acl{Defiso}
\citeman{http://www.cs.utexas.edu/users/moore/acl2/manuals/current/manual/?topic=ACL2____DEFISO}{defiso}
takes four functions as (some of its) inputs:
\begin{itemize}[nosep]
\item
\acl{old}, the predicate for the old data representation,
like $X$ or $Y$ in \secref{sec:overview}.
\item
\acl{new}, the predicate for the new data representation,
like $X'$ or $Y'$ in \secref{sec:overview}.
\item
\acl{iso}, the isomorphism from \acl{old} to \acl{new},
like $\isoin$ or $\isoout$ in \secref{sec:overview}.
\item
\acl{osi}, the isomorphism from \acl{new} to \acl{old},
like $\osiin$ or $\osiout$ in \secref{sec:overview}.
\end{itemize}

\acl{Defiso} attempts to prove the following applicability conditions
(assuming unary functions):%
\footnote{Sometimes this paper uses infix notations
for logical connectives and equality.}
\begin{itemize}[nosep]
\item
$\acl{(old o)}\ \Longrightarrow\ \acl{(new (iso o))}$,
i.e.\ \acl{iso} maps values in \acl{old} to values in \acl{new}.
\item
$\acl{(new n)}\ \Longrightarrow\ \acl{(old (osi n))}$,
i.e.\ \acl{osi} maps values in \acl{new} to values in \acl{old}.
\item
$\acl{(old o)}\ \Longrightarrow\ \acl{(osi (iso o))}\ =\ \acl{o}$,
i.e.\ \acl{osi} is \acl{iso}'s inverse over the values in \acl{old}.
\item
$\acl{(new n)}\ \Longrightarrow\ \acl{(iso (osi n))}\ =\ \acl{n}$,
i.e.\ \acl{iso} is \acl{osi}'s inverse over the values in \acl{new}.
\end{itemize}
As in APT transformations,
it is the user's responsibility to ensure that these conditions are proved,
possibly by supplying hints to \acl{defiso}
or by proving lemmas before calling \acl{defiso}.

If these proofs are successful,
\acl{defiso} generates theorems asserting
the injectivity of \acl{iso} and \acl{osi} over the predicates:
\begin{itemize}[nosep]
\item
$\acl{(old o1)}\ \wedge\ \acl{(old o2)} \ \Longrightarrow\
 [\acl{(iso o1)}\ =\ \acl{(iso o2)}\ \Longleftrightarrow\
  \acl{o1}\ =\ \acl{o2}]$
\item
$\acl{(new n1)}\ \wedge\ \acl{(new n2)}\ \Longrightarrow\
 [\acl{(osi n1)}\ =\ \acl{(osi n2)}\ \Longleftrightarrow\
  \acl{n1}\ =\ \acl{n2}]$
\end{itemize}
The proof hints for these injectivity theorems
are generated automatically from the applicability conditions
\cite{defiso-design-notes},%
\footnote{If $\phi$ has a left inverse $\funinv{\phi}$,
then $\phi$ is injective:
assuming $\phi(z_1) = \phi(z_2)$,
apply $\funinv{\phi}$ to both sides
to obtain $\funinv{\phi}(\phi(z_1)) = \funinv{\phi}(\phi(z_2))$,
and use the left inverse property to conclude $z_1 = z_2$.}
similarly to how the APT transformations
generate $\refsto$ proofs from the applicability conditions
(see \secref{sec:intro}).

Information about
all the above functions, applicability conditions, and theorems
is stored into an ACL2 table,
under a name supplied as another input to \acl{defiso}.
In the call \acl{(defiso isomap old new iso osi ...)},
\acl{isomap} is such a name,
and \acl{...} consists of hints and other optional inputs.

\acl{Defiso} optionally, and by default,
generates additional applicability conditions
about the guards of the predicates and isomorphisms:
\begin{itemize}[nosep]
\item
The guards of \acl{old} and \acl{new} are always true,
so that \acl{old} and \acl{new} can be applied to any value.
\item
The guard of \acl{iso} includes \acl{old},
so that \acl{iso} can be applied to any value in \acl{old}.
\item
The guard of \acl{osi} includes \acl{new},
so that \acl{osi} can be applied to any value in \acl{new}.
\end{itemize}
These ensure that certain terms involving the predicates and isomorphisms
are guard-verified \cite{defiso-design-notes}.

The predicates may take multiple arguments,
whose number must suitably match
the number of arguments and results of the isomorphism;
two or more results are returned via \acl{mv}
\citeman{file:///Users/ac/Work/acl2/kepo/doc/manual/index.html?topic=ACL2____MV}{mv}.
Thus, the old and new data representations
may be subsets of cartesian products of the universe of ACL2 values.
In this case, the applicability conditions involve \acl{mv-let}
\citeman{http://www.cs.utexas.edu/users/moore/acl2/manuals/current/manual/?topic=ACL2____MV-LET}{mv-let}.

In addition to function symbols,
untranslated lambda expressions and macro symbols
can also be supplied to \acl{defiso}
to specify the predicates and isomorphisms.
A macro symbol \acl{mac} abbreviates its eta expansion
\acl{(lambda (z1 z2 ...) (mac z1 z2 ...))},
limited to just the macro's required arguments \acl{z1}, \acl{z2}, etc.

While \acl{defiso} was developed to support isomorphic transformations,
it may have more general uses.
It automates the formulation of isomorphism theorems,
and the proof of injectivity theorems from them,
for predicates and functions
that are not necessarily used in isomorphic transformations.


\subsection{Initiating Isomorphic Transformations}
\label{sec:isodata}

Consider an isomorphic mapping \acl{(defiso isomap old new iso osi ...)},
like $\xymatrix@1{X\ar[r]|{\;\isoin\;} & X'}$
or $\xymatrix@1{Y\ar[r]|{\;\isoout\;} & Y'}$
in \secref{sec:overview},
for changing the representation of the inputs and outputs
of an ACL2 function of the form
\begin{bacl}
  (defun f (x)
    (declare (xargs :guard (g x) :measure (m x)))
    (if (a x) (b x) (c x (f (d x)))))
\end{bacl}
which is like $f$ in \secref{sec:overview}
with the addition of a guard and an explicit measure
(like $\meas$ in \secref{sec:overview}),
assuming that $\xymatrix@1{X\ar[r]|{\;\isoin\;} & X'}$
and $\xymatrix@1{Y\ar[r]|{\;\isoout\;} & Y'}$ are the same
(i.e.\ $X = Y$, $X' = Y'$, etc.).

\acl{Isodata}
\citeman{http://www.cs.utexas.edu/users/moore/acl2/manuals/current/manual/?topic=APT____ISODATA}{isodata}
takes \acl{f} and \acl{isomap} as (some of its) inputs,
and attempts to prove the following applicability conditions:%
\footnote{The applicability conditions for \acl{(defiso isomap ...)}
can be regarded as additional, ``indirect'' applicability conditions
of the \acl{isodata} call.}
\begin{itemize}[nosep]
\item
$\acl{(old x)}\ \Longrightarrow\ \acl{(old (f x))}$,
i.e.\ \acl{f} maps inputs with the old representation
to outputs with the old representation.
\item
$\acl{(old x)}\ \wedge\ \neg\acl{(a x)}\ \Longrightarrow\ \acl{(old (d x))}$,
i.e.\ the old representation is preserved across recursive calls.
\item
$\acl{(g o)}\ \Longrightarrow\ \acl{(old o)}$,
i.e.\ the inputs in the guard of \acl{f}
(i.e.\ the ones over which \acl{f} is ``well-defined'')
all have the old representation.
This is an optional applicability condition,
present when guards are to be verified (which is the default).
\end{itemize}

If these proofs (which are the user's responsibility) succeed,
\acl{isodata} generates a new function
\begin{bacl}
  (defun f1 (x)
    (declare (xargs :guard (and (new x) (g (osi x))) :measure (m (osi x))))
    (if (mbt$ (new x))
        (if (a (osi x)) (iso (b (osi x))) (iso (c (osi x) (osi (f1 (iso (d (osi x))))))))
      nil)) ; irrelevant
\end{bacl}
which is like $f'$ in \secref{sec:overview},
with the addition of a guard and a wrapping \acl{if},
both explained next.

The guard's first conjunct \acl{(new x)} ensures that:
(i) the call \acl{(osi x)} in the second conjunct is guard-verified
(see the third optional applicability condition of \acl{defiso}
in \secref{sec:defiso});
and (ii) \acl{f1} only operates (according to its guard)
on inputs in the new representation,
analogously to how \acl{f} only operates on inputs in the old representation
(see the third applicability condition of \acl{isodata} above).%
\footnote{The guard's second conjunct alone
does not prevent an \acl{x} outside \acl{new}
to be mapped to an \acl{(osi x)} in \acl{old}.}
The guard's second conjunct \acl{(g (osi x))} further ensures that
\acl{f1} only operates on the subset of the new representation
isomorphic to the guard of \acl{f}.
The guard proof hints, if guards are to be verified,
are generated automatically from the applicability conditions
\cite{isodata-design-notes}.

The wrapping \acl{(if (mbt\$ (new x)) ...)},
where \acl{mbt\$}
\citeman{http://www.cs.utexas.edu/users/moore/acl2/manuals/current/manual/?topic=ACL2____MBT_42}{mbt\$}
is a variant of \acl{mbt} that requires non-\acl{nil} instead of \acl{t}
(to avoid requiring \acl{new} to be boolean-valued),
is needed because the ACL2 language is untyped
(unlike the ``mathematical language'' in \secref{sec:overview}).
Without it, the termination of \acl{f1} would not be always provable
from the termination of \acl{f} with the measure \acl{(m (osi x))}.
In this case \acl{(mbt\$ (new x))} could be negated
and disjoined before \acl{(a (osi x))},
but the form above is the most general,
and the branch marked as `irrelevant' can be anything;
this branch always trivially passes guard verification,
because it is unreachable under the guard.
The termination proof hints
are generated automatically from the applicability conditions
and the termination theorem of \acl{f}
\cite{isodata-design-notes}.

\acl{Isodata} also generates the following theorems,
whose proof hints are generated automatically
from the applicability conditions \cite{isodata-design-notes}:
\begin{itemize}[nosep]
\item
$\acl{(old x)}\ \Longrightarrow\ \acl{(f x)}\ =\ \acl{(osi (f1 (iso x)))}$,
which is like
$f = \funcompcomp{\osiout}{f'}{\isoin}$ in \secref{sec:overview},
with the hypothesis added because the ACL2 language is untyped.
\item
$\acl{(new x)}\ \Longrightarrow\ \acl{(f1 x)}\ =\ \acl{(iso (f (osi x)))}$,
which is like
$f' = \funcompcomp{\isoout}{f}{\osiin}$ in \secref{sec:overview},
with the hypothesis added because the ACL2 language is untyped.
\item
$\acl{(new x)}\ \Longrightarrow\ \acl{(new (f1 x))}$,
i.e.\ \acl{f1} maps inputs with the new representation
to outputs with the new representation,
analogously to the first applicability condition of \acl{isodata} above.
\end{itemize}
We also generate an incompatibility theory invariant
\citeman{http://www.cs.utexas.edu/users/moore/acl2/manuals/current/manual/?topic=ACL2____THEORY-INVARIANT}{theory-invariant}
preventing the first and second theorems from being simultaneously enabled.

The above is accomplished via the call
\acl{(isodata f (((x :result) isomap)) ...)},
where the doublet \acl{((x :result) isomap)} specifies
the transformation of the argument \acl{x} and of the  result via \acl{isomap},
and \acl{...} consists of hints and other options.
The paper's supporting materials%
\footnote{File
\acl{[books]/workshops/2020/coglio-westfold/schematic-example.lisp}.}
includes a development of this ``schematic'' example.

\acl{Isodata} supports non-recursive functions,
as well as recursive functions with multiple base cases and/or recursive calls,
but not yet mutually recursive functions.
It supports the transformation of each argument and result
with a possibly different isomorphic mapping:
the second input of \acl{isodata} is a list of doublets
\acl{((arg/res-list1 isomap1) (arg/res-list2 isomap2) ...)}
that specifies that the arguments and results listed in \acl{arg/res-listk}
are transformed according to the isomorphic mapping \acl{isomapk};
results of multi-valued functions
are denoted via \acl{:result1}, \acl{:result2}, etc.
Isomorphic mappings involving tuples of arguments or results
are not yet supported.
Instead of referring to existing \acl{defiso}s,
it is possible to specify isomorphic mappings ``inline'',
e.g.\ \acl{(... (arg/res-list (old new iso osi)) ...)},
in which case \acl{isodata}
internally generates and uses local \acl{defiso}s.

\acl{Isodata} also supports a \acl{:predicate} option, \acl{nil} by default.
When set to \acl{t}, this option specifies that the target function \acl{old}
is (treated like) a predicate,
and \acl{isodata} generates
slightly different applicability conditions and results in this case.
See the documentation
\citeman{http://www.cs.utexas.edu/users/moore/acl2/manuals/current/manual/?topic=APT____ISODATA}{isodata}
for details.

After obtaining a function like \acl{f1} above,
the user can use APT's \acl{simplify} transformation \cite{simplify}
to expand the definitions of the isomorphisms
and rephrase the computations
to operate directly on the new input and output representations,
without any remaining references to
the isomorphisms and the old representations.
This process is user-guided in general.


\subsection{Propagating Isomorphic Transformations}
\label{sec:propiso}

The automated process of propagating isomorphic transformations of a set of interface
functions for a data type to all the higher-level functions that use them is performed by
the \acl{propagate-iso} transformation. 


The first task of \acl{propagate-iso} is to find all the functions and theorems that
reference the interface functions directly or indirectly. This is done by traversing the
event history starting from the definitions of the interface functions.

{\bf Type Inference.}
In order to generate the isomorphism theorems relating the old functions to their
isomorphic versions (as in \secref{sec:isodata}) we need to know which arguments and
results have the type being transformed. As ACL2 is untyped, \acl{propagate-iso} needs to
do this type inference itself. The types of arguments are found by simple traversal of the
guards of the functions, and the result types are found by looking at the bodies of
functions and also by looking for theorems that specify them.

{\bf Generating Dependent Isomorphisms.}
Frequently the data type being isomorphically transformed is a component of a larger data
type such as a record, list, or map. The dependent isomorphism can be constructed by
mapping each element of the larger data type using the base isomorphism if it is of the
base type, otherwise using identity, which is a trivial isomorphism. \acl{Propagate-iso}
tries to construct a dependent isomorphism whenever it encounters a predicate that uses
the isomorphism predicate. It uses the structure of the predicate to generate the
isomorphism functions.

For example, if we have the isomorphism \acl{(defiso isomap old new iso osi ...)} and a
predicate that is a pair of \acl{natp} and \acl{old}
\begin{bacl}
  (defun P (x) (and (natp (car x)) (old (cdr x))))
\end{bacl}
we get the isomorphic predicate and conversion functions
\begin{bacl}
  (defun P-new (x) (and (natp (car x)) (new (cdr x))))
  (defun P-to-P-new (x) (cons (car x) (iso (cdr x))))
  (defun P-new-to-P (x) (cons (car x) (osi (cdr x))))
\end{bacl}

An example of the dependent isomorphism for a list of \acl{old} is given in \secref{sec:dr-plan}.

{\bf Generated Theorems and Proof Hints.}
All of the functions and theorems produced by
\acl{propagate\-iso} must be accepted by ACL2. This means the guard and termination
conditions as well as the theorems must be proved by ACL2. We can give a meta-level
argument that all these must be true, but ACL2 has to prove them at the object level.  To
generate hints for the ACL2 proofs, \acl{propagate-iso} maintains three rulesets
\citeman{http://www.cs.utexas.edu/users/moore/acl2/manuals/current/manual/?topic=RULESETS}{rulesets}:
one for old-to-new rules, one for new-to-old rules, and one for generally-useful rules,
mainly typing rules and the \acl{defiso} theorems, with a few list data-structure rules.

For the transformed definitions and theorems, transformed proofs can be performed in
principle. ACL2 does not retain proofs so this is not possible, but it can be approximated
by use of transformed hints. This works when the hints are explicit, but does not take
account of the proving environment, in particular, whether rules have been enabled or
disabled externally since introduction.

An alternative to reproducing the original proof is to prove the new theorem from the old
theorem. To do this \acl{propagate-iso} generates a hint with a \acl{:use} of the old
theorem instantiated with the necessary new-to-old conversions, and a theory
consisting of the general ruleset and the new-to-old ruleset.

Because these proof hint strategies are not guaranteed to succeed, \acl{propagate-iso}
tries both approaches: it gives instructions to first try to prove using the
transformed hints and if that does not succeed, try to prove using the old theorem. As a
last resort it tries proving with the current theory.

To prove the isomorphism theorems for a new introduced function, first the new-to-old
theorem is proved. This is necessary in general because all the theorems on the old
function are available, but the corresponding theorems on the new function will not be
introduced until after the two isomorphism theorems. The rules for mapping
previously-defined new functions into old functions have few preconditions so occurrences
of new functions are transformed into expressions involving old functions, and then the
old theorems can be used to complete the proofs. The old-to-new theorem follows
easily from the new-to-old theorem and the isomorphism theorems.

As the proof hints generated by \acl{propagate-iso} are not alway guaranteed to succeed,
syntax is provided to allow the user to augment or override the generated hints for any
theorem. At present, none of our examples require this level of user intervention.


\section{Examples}
\label{sec:examples}


\subsection{Unbounded and Bounded Integers}

Popular programming languages like C and Java
typically use bounded integer types and operations,
while requirements specifications
typically use unbounded integer types and operations.
Thus, synthesizing a C or Java program from a specification,
or proving that a C or Java program complies with a specification,
often involves showing that unbounded and bounded integers are ``equivalent''
under the preconditions stated by the specification.

When using APT as outlined in \secref{sec:intro}
for these synthesis or verification tasks,
showing that equivalence amounts to
transforming between unbounded and bounded types and operations.
Previous work \cite{simplify}
exemplified a general methodology to transform
bounded integer operations into unbounded integer operations via rewriting;
it also alluded at isomorphic data type transformations
for transforming bounded integer types into unbounded integer types.
An unbounded type is not isomorphic to a bounded type for cardinality reasons,
but a finite subset of the former may be isomorphic to the latter.
When the representation of the bounded type is not just a range of integers
(see examples below),
there is a non-trivial isomorphism between
the bounded type and the corresponding subset of the unbounded type.

The aforementioned previous work \cite{simplify} uses, as an example,
this Java implementation of Bresenham's line drawing algorithm
(whose details are completely unimportant here):
\begin{bacl}
  // draw a line from (0, 0) to (a, b), for a and b below a given limit:
  static void drawLine(int a, int b) \{
      int x = 0, y = 0, d = 2 * b - a;
      while (x <= a) \{
          drawPoint(x, y); // on screen
          x++;
          if (d >= 0) \{ y++; d += 2 * (b - a); \}
          else \{ d += 2 * b; \}
      \}
  \}
\end{bacl}
The ACL2 representation of the Java code (see the cited paper for details)
has the form
\begin{bacl}
  (defun drawline-loop (a b x y d screen) ; loop of the method
    (if (invariant a b x y d)
        (if (not (lte32 x a))
            screen ; exit loop
          (drawline-loop a b
                         (add32 x (int32 1))
                         (if (gte32 d (int32 0))
                             (add32 y (int32 1))
                           y)
                         (if (gte32 d (int32 0))
                             (add32 d (mul32 (int32 2) (sub32 b a)))
                           (add32 d (mul32 (int32 2) b)))
                         (drawpoint x y screen)))
      :undefined))
  (defun drawline (a b screen) ; method
    (if (precondition a b)
        (drawline-loop a b
                       (int32 0) ; x
                       (int32 0) ; y
                       (sub32 (mul32 (int32 2) b) a) ; d
                       screen)
      :undefined))
\end{bacl}
where:
\acl{add32}, \acl{sub32}, \acl{mul32}, \acl{lte32}, and \acl{gte32}
are 32-bit signed integer operations;
and \acl{int32} converts from \acl{(lambda (x) (signed-byte-p 32 x))}
to a type of unbounded integers recognized by \acl{int32p}.
For the purpose of this example,
the exact definitions of \acl{int32p}, \acl{int32}, \acl{add32}, etc.\
are unimportant---%
the paper's supporting materials%
\footnote{File
\acl{[books]/workshops/2020/coglio-westfold/integer-example.lisp}.}
introduce them as constrained functions.
For concreteness, they could be
\acl{int-value-p} \citeman{http://www.cs.utexas.edu/users/moore/acl2/manuals/current/manual/?topic=JAVA____INT-VALUE-P}{int-value-p},
\acl{make-int-value} \citeman{http://www.cs.utexas.edu/users/moore/acl2/manuals/current/manual/?topic=JAVA____MAKE-INT-VALUE}{make-int-value},
\acl{int-add} \citeman{http://www.cs.utexas.edu/users/moore/acl2/manuals/current/manual/?topic=JAVA____INT-ADD}{int-add},
etc.\
in the formalization of Java primitive values and operations
in \acl{[books]/kestrel/} \acl{java/language};
or they could be
\acl{(lambda (x) (unsigned-byte-p 32 x))},
\acl{(lambda (x) (loghead 32 x))},
\acl{(lambda (x y) (bvplus 32 x y))}, etc.\
in the bit vector library in \acl{[books]/kestrel/bv}.

Consider the isomorphic mapping
\begin{bacl}
  (defiso isomap ; name
          int32p ; old representation
          (lambda (x) (signed-byte-p 32 x)) ; new representation
          int ; conversion from old to new representation
          int32) ; conversion from new to old representation
\end{bacl}
where \acl{int} converts from \acl{int32p}
to \acl{(lambda (x) (signed-byte-p 32 x))};
again, the exact definition of \acl{int} is unimportant,
but for concreteness it could be \acl{int-value->int}
\citeman{http://www.cs.utexas.edu/users/moore/acl2/manuals/current/manual/?topic=JAVA____INT-VALUE}{int-value}
in the formalization of Java primitive values and operations
in \acl{[books]/kestrel/java/language},
or \acl{(lambda (x) (logext 32 x))}
in the bit vector library in \acl{[books]/kestrel/bv}.

Applying \acl{isodata}
to \acl{drawline-loop} and \acl{drawline} with \acl{isomap},
changes the representation of \acl{a}, \acl{b}, \acl{x}, \acl{y}, and \acl{d}
from \acl{int32p} to
\acl{(lambda (x) (signed-byte-p 32 x))}.
This puts \acl{int32} in front of each occurrence of \acl{a}, \acl{b}, etc.,
and puts \acl{int} in front of each recursive call argument,
which produces some terms of the form \acl{(int (int32 ...))}:
\begin{bacl}
  (defun drawline-loop1 (a b x y d screen)
    ...
    (if (not (lte32 (int32 x) (int32 a)))
        screen
      (drawline-loop1 (int (int32 a))
                      (int (int32 b))
                      (int (add32 (int32 x) (int32 1)))
                      (int (if (gte32 (int32 d) (int32 0))
                               (add32 (int32 y) (int32 1))
                             (int32 y)))
                      (int (if (gte32 (int32 d) (int32 0))
                               (add32 (int32 d)
                                      (mul32 (int32 2)
                                             (sub32 (int32 b) (int32 a))))
                             (add32 (int32 d)
                                    (mul32 (int32 2) (int32 b)))))
                      (drawpoint (int32 x) (int32 y) screen)))
    ...)
  (defun drawline1 (a b screen)
    ...
    (drawline-loop1 (int32 a) (int32 b) (int32 0) (int32 0)
                    (sub32 (mul32 (int32 2) (int32 b))
                           (int32 a))
                    screen)
    ...)
\end{bacl}

Then applying APT's \acl{simplify} \cite{simplify}
with the rewrite rules for transforming \acl{add32} to \acl{+} etc.,
which temporarily produces further terms of the form \acl{(int (int32 ...))},
results in the following functions,
whose bodies no longer use the bounded type and operations
(except for the call in \acl{drawpoint},
but that could be also similarly transformed):
\begin{bacl}
  (defun drawline-loop2 (a b x y d screen)
    ...
    (if (< a x)
        screen
      (drawline-loop2 a b
                      (+ 1 x)
                      (if (< d 0) y (+ 1 y))
                      (if (< d 0)
                          (+ d (* 2 b))
                        (+ d (- (* 2 a)) (* 2 b)))
                      (drawpoint (int32 x) (int32 y) screen)))
    ...)
  (defun drawline2 (a b screen)
    ...
    (drawline-loop2 a b 0 0 (+ (- a) (* 2 b)) screen)
    ...)
\end{bacl}

The technique exemplified above extends
the technique exemplified in previous work \cite{simplify},
from transforming just bounded integer operations into unbounded ones,
to transforming both bounded integer types and bound integer operations
into unbounded ones.


\subsection{Loop Re-Indexing}
\label{sec:loop}

A perhaps less expected example of isomorphic data type transformation
is the re-indexing of a loop to count up instead of down (or vice versa).
In this case, the old and new representation are the same,
but the isomorphic conversions ``flip'' the loop range.

As a simple but suggestive example,
consider the tenfold application of a unary function \acl{h}:%
\footnote{Fixed-length loops are common in cryptography, for instance.}
\begin{bacl}
  (defun applyn (x n) ; apply h to x for n times -- (h (h (h ... (h x)...)))
    (declare (xargs :guard (and (natp n) (<= n 10))))
    (if (zp n) x (h (applyn x (1- n)))))
  (defun applyten (x) ; apply h to x for 10 times
    (declare (xargs :guard t))
    (applyn x 10))
\end{bacl}

The two functions above could be part of
a requirements or intermediate specification.
Presumably a requirements specification
would omit the condition \acl{(<= n 10)} from \acl{applyn},
but that condition could be added by the \acl{restrict} transformation
\citeman{http://www.cs.utexas.edu/users/moore/acl2/manuals/current/manual/?topic=APT____RESTRICT}{restrict}
as a preparatory refinement step prior to the ones described below,
given that \acl{applyten} calls \acl{applyn} with 10 as \acl{n}:
this way, \acl{applyn} above would be part of an intermediate specification.
Regardless, \acl{applyn} exhibits a simple recursion on the natural numbers
with measure and termination proof easily found by ACL2.
This ``loop'' counts down, from 10 to 0,
which is natural in a functional program or specification.

The set $\{0,\ldots,10\}$ of the possible values of the loop variable \acl{n}
is isomorphic to itself in more than one way
(i.e.\ the obvious way, where the isomorphisms are identity).
In particular,
the function that maps each element $n\in\{0,\ldots,10\}$ to $10-n$
is an isomorphism over the set, with itself as the inverse.
The following isomorphic transformation step,
where the \acl{defiso} is generated on the fly, succeeds:
\begin{bacl}
  (isodata applyn ((n ((lambda (n) (and (natp n) (<= n 10)))
                       (lambda (n) (and (natp n) (<= n 10)))
                       (lambda (n) (- 10 n))
                       (lambda (n) (- 10 n)))))
           :new-name applyn0)
\end{bacl}
The resulting function includes arithmetic expressions
such as \acl{(+ 10 (- (+ -1 10 (- n))))},
amenable to simplification
via APT's \acl{simplify} transformation \cite{simplify},
which produces
\begin{bacl}
  (defun applyn1 (x n)
    (declare (xargs :guard ... :measure (acl2-count (+ 10 (- n)))))
    (if (and (natp n) (<= n 10))
        (if (< n 10)
            (h (applyn1 x (+ 1 n)))
          x)
      nil))
\end{bacl}
Now the loop counts up, from 0 to 10,
which is more complicated in a functional program or specification
(see the measure of \acl{applyn1}),
but more natural or common in an imperative implementation,
which is the direction where this loop re-indexing transformation is headed.

See the paper's supporting materials%
\footnote{File \acl{[books]/workshops/2020/coglio-westfold/loop-example.lisp}.}
for a complete development of this example.

For this simple example, one could manually write
the new functions, and the theorems that relate them to the old functions,
without any additional proof effort: ACL2 proves the theorems automatically.
However, calling \acl{isodata} and \acl{simplify} seems easier,
and explicates the principles that lead from the old loop to the new loop.
With more complex examples,
the manually written theorems may no longer be proved automatically,
perhaps due to ``interference'' from other rules
that apply to other parts of the functions.
In contrast, \acl{isodata} is designed to generate
proof hints for termination, guards, and theorems
that are precisely targeted to the proof goals
(e.g.\ that do not depend on the current theory)
and that are expected to always succeed.
With reference to \secref{sec:intro},
manually transforming a loop amounts to
writing $\spec_{i+1}$ and verifying $\spec_i \refsto \spec_{i+1}$,
while using automated transformations amounts to
generating $\spec_{i+1}$ and the proof of $\spec_i \refsto \spec_{i+1}$
from $\spec_i$.


\subsection{Efficient Value Caching With Invariant Maintenance}
\label{sec:dr-plan}

Our largest example is a drone route planner. The problem is to have a set of drones visit
a set of sites. This is a distributed system with route planning interleaved with
execution. At each step each drone generates candidate plans for where to visit
next. These plans are sent to a coordinator that filters these candidate plans to reduce
redundancy among the different drones. Each drone then chooses one of the filtered plans
to execute. After execution of the (partial) plan, the plan-coordinate-execute cycle is
repeated until all the target locations have been visited.

The key data type is the state of a drone, \acl{drone-st}. A list of drone states, one for
each drone, is passed around the high-level functions. The drone state has four fields:
the identifier for the drone; a graph which gives the possible moves from each location
(\acl{node}); the nodes that have been visited; and the path that has been taken so far in
the planning process:\footnote{The use of the FTY
\citeman{http://www.cs.utexas.edu/users/moore/acl2/manuals/current/manual/?topic=ACL2____FTY}{fty}
package here is not essential. Any product/record constructs could be used.}
\begin{bacl}
  (fty::defprod dr-state
    ((drone-id drone-id)
     (dgraph dgraph-p)
     (visited-nodes node-list)
     (path-taken node-list)))
\end{bacl}

During the planning, it is important to know the current location of each node and the
nodes that are unvisited at each point in the planning process. A natural optimization
is to store these values in the drone state record and maintain them incrementally instead
of recomputing them from the path taken by the drone and its set of visited nodes.

{\bf Isomorphism Definition.}
The first step of this optimization is to define a new state record with fields for the
values to be maintained:
\begin{bacl}
  (fty::defprod dr-state-ext0
    ((drone-id drone-id)
     (dgraph dgraph-p)
     (visited-nodes node-list)
     (path-taken node-list)
     (unvisited-nodes node-list)
     (currentpos node-or-null)))
\end{bacl}

We define functions to map back and forth between these two types: \acl{from-dr-state-ext}, which just
omits the two new fields; and \acl{to-dr-state-ext} which computes the values of the two
new fields. We also define theorems to relate the accessor functions of the two types
using these functions. For example:
\begin{bacl}
  (defthm dr-state->drone-id-~>dr-state-ext0->drone-id
    (equal (dr-state->drone-id drn-st)
           (dr-state-ext0->drone-id (to-dr-state-ext drn-st))))
\end{bacl}

To define a predicate representing a type isomorphic to \acl{dr-state}, we restrict the
two new fields to be functions of the other fields:
\begin{bacl}
  (define dr-state-ext-p (drn-st)
    :returns (b booleanp)
    (and (dr-state-ext0-p drn-st)
         (equal (dr-state-ext0->unvisited-nodes drn-st)
                (set-difference-equal (dgraph->nodes (dr-state-ext0->dgraph drn-st))
                                      (dr-state-ext0->visited-nodes drn-st)))
         (equal (dr-state-ext0->currentpos drn-st)
                (drone-location (from-dr-state-ext drn-st)))))
\end{bacl}

The isomorphism is then:
\begin{bacl}
  (defiso dr-state-p-to-dr-state-ext-p
    dr-state-p dr-state-ext-p to-dr-state-ext from-dr-state-ext
    :hints (:beta-of-alpha (("Goal" :in-theory (enable from-dr-state-ext to-dr-state-ext)))
              :alpha-of-beta (("Goal" :in-theory (enable from-dr-state-ext to-dr-state-ext)))))
\end{bacl}

Given the basic isomorphism we are now ready to perform the isomorphic type refinement.
Apart from the accessor functions, \acl{extend-path-taken} is the only function that
accesses the internals of \acl{dr-state}. It is the function that incorporates a new plan
into the state, extending the \acl{path-taken} and adding the plan nodes to the
\acl{visited-nodes}.
\begin{bacl}
  (define extend-path-taken ((drn-st good-dr-state-p) (plan node-list-p))
    :guard (or (null plan) (valid-plan-p (drone-location drn-st) plan drn-st))
    (change-dr-state drn-st
                    :path-taken (append (dr-state->path-taken drn-st) plan)
                    :visited-nodes (union-equal (dr-state->visited-nodes drn-st)
                                                plan)))
\end{bacl}

{\bf Initial Propagation.}  The guard uses three functions that depend on \acl{dr-state}:
\acl{good-dr-state-p}, \acl{valid-plan-p} and \acl{drone-location}, where
\acl{good-dr-state-p} and \acl{valid-plan-p} are simple well-formedness predicates on
\acl{dr-state} and \acl{plan} types respectively. We want isomorphic versions of these
three functions before transforming \acl{extend-path-taken}. As these functions do not
access \acl{dr-state} directly, their isomorphic versions can be generated using
\acl{propagate-iso}:
\begin{bacl}
  (propagate-iso dr-state-p-to-dr-state-ext-p
                 ((dr-state dr-state-ext  ; Constructors
                   dr-state-~>dr-state-ext dr-state-ext-~>dr-state
                   (* * * *) => (dr-state-p))
                  (dr-state->drone-id dr-state-ext0->drone-id   ; Destructors
                   dr-state->drone-id-~>dr-state-ext0->drone-id
                   dr-state-ext0->drone-id-~>dr-state->drone-id
                   (dr-state-p) => *)
                   ...
                 --Similar entries for the other 3 destructors--
                 )
                 :first-event dr-state*-p
                 :last-event valid-plan-p-node-path-p)
\end{bacl}

The first argument \acl{dr-state-p-to-dr-state-ext-p} is the name of the isomorphism,
which could in general be a list of isomorphisms to be applied in parallel. Then follows
information about the interface functions, in this case just the constructor and
destructor functions. For example, the first element specifies that for the constructor
function \acl{dr-state}, \acl{dr-state-ext} is the isomorphic constructor function,
\acl{dr-state-\textasciitilde>dr-state-ext} and
\acl{dr-state-ext-\textasciitilde>dr-state} are the theorems that can
transform one into the other, and \acl{(* * * *) => (dr-state-p)} is the signature of
\acl{dr-state}. This signature specifies that \acl{dr-state} takes four arguments not of
any isomorphism type and returns a single value satisfying the \acl{dr-state-p} predicate.
The \acl{:first-event} and \acl{:last-event} specify the range of events for which to consider
creating isomorphic versions.

{\bf Dependent Isomorphisms.}
For predicates that include a call to an existing isomorphism predicate, \acl{propagate-iso} not only
creates an isomorphic version of the predicate, but also constructs an isomorphism between the
two. For example, for the predicate \acl{all-dr-state-p} defined as
\begin{bacl}
  (defun all-dr-state-p (drn-sts)
    (if (atom drn-sts)
        (null drn-sts)
      (and (dr-state-p (first drn-sts))
           (all-dr-state-p (rest drn-sts)))))
\end{bacl}
it defines the isomorphic predicate \acl{all-dr-state-ext-p} as
\begin{bacl}
  (defun all-dr-state-ext-p (drn-sts)
    (if (atom drn-sts)
        (null drn-sts)
      (and (dr-state-ext-p (first drn-sts))
           (all-dr-state-ext-p (rest drn-sts)))))
\end{bacl}
and, based on the structure of the predicate, creates the conversion functions
\begin{bacl}
  (defun all-dr-state-p-->-all-dr-state-ext-p (drn-sts)
    (if (atom drn-sts)
        nil
      (cons (to-dr-state-ext (first drn-sts))
            (all-dr-state-p-->-all-dr-state-ext-p (rest drn-sts)))))
  (defun all-dr-state-ext-p-->-all-dr-state-p (drn-sts)
    (if (atom drn-sts)
        nil
      (cons (from-dr-state-ext (first drn-sts))
            (all-dr-state-ext-p-->-all-dr-state-p (rest drn-sts)))))
\end{bacl}
and creates the isomorphism
\begin{bacl}
  (defiso all-dr-state-p-iso-all-dr-state-ext-p
    all-dr-state-p all-dr-state-ext-p
    all-dr-state-p-->-all-dr-state-ext-p all-dr-state-ext-p-->-all-dr-state-p
    :hints ...)
\end{bacl}

{\bf Last Initialization.}
After \acl{propagate-iso} has generated isomorphic versions of the three functions used
in its guard, the isomorphic version of \acl{extend-path-taken}
is generated using \acl{isodata} and optimized by \acl{simplify}:
\begin{bacl}
  (isodata extend-path-taken (((drn-st :result) good-dr-state-p-iso-good-dr-state-ext-p)))
\end{bacl}
which specifies that the \acl{drn-st} argument and the result are of the isomorphism
source type, and generates the isomorphic version of
\acl{extend-path-taken}:\footnote{Currently \acl{isodata} and \acl{simplify} produce
  results that are equivalent to these but a little more syntactically complex, using
  \acl{defun}, \acl{mbt} and extra let-binding. The simplified form presented here is
  intended to facilitate comparison with the original definition of
  \acl{extend-path-taken}. Also, the guards are omitted for brevity.}
\begin{bacl}
  (define extend-path-taken$1 ((drn-st good-dr-state-ext-p) (plan node-list-p))
    (to-dr-state-ext (let ((new-drn-st (from-dr-state-ext drn-st)))
                       (dr-state (dr-state->drone-id new-drn-st)
                                 (dr-state->dgraph new-drn-st)
                                 (union-equal (dr-state->visited-nodes new-drn-st)
                                              plan)
                                 (append (dr-state->path-taken new-drn-st)
                                         plan)))))
\end{bacl}
Then the \acl{simplify} form:
\begin{bacl}
  (simplify extend-path-taken$1
            :assumptions ((good-dr-state-ext-p drn-st))
            :enable (dr-state-ext set-difference-equal-2-append-rev))
\end{bacl}
produces the optimized version:
\begin{bacl}
  (define extend-path-taken$2 ((drn-st good-dr-state-ext-p) (plan node-list-p))
    (dr-state-ext0 (dr-state-ext0->drone-id drn-st)
                   (dr-state-ext0->dgraph drn-st)
                   (union-equal (dr-state-ext0->visited-nodes drn-st) plan)
                   (append (dr-state-ext0->path-taken drn-st) plan)
                   (set-difference-equal (dr-state-ext0->unvisited-nodes drn-st)
                                         plan)
                   (if (consp plan)
                       (car (last plan))
                     (dr-state-ext0->currentpos drn-st))))
\end{bacl}
\acl{Simplify} not only eliminates the conversion functions \acl{to-dr-state-ext} and
\acl{from-dr-state-ext}, but the enabling of the interface function \acl{dr-state-ext}
and specific distributive rules allow the
\acl{unvisited\-nodes}
and \acl{currentpos} fields
to be computed incrementally rather than using the relatively expensive expressions in the
definition of \acl{dr-state-ext-p} above.

\acl{Isodata} produces theorems relating \acl{extend-path-taken} and
\acl{extend-path-taken\$1}; \acl{simplify} produces theorems relating
\acl{extend-path-taken\$1} and \acl{extend-path-taken\$2}. However \acl{propagate\-iso}
requires theorems relating \acl{extend-path-taken} and \acl{extend-path-taken\$2}, 
which are proved by simply chaining the \acl{isodata} and \acl{simplify} theorems.

{\bf Final Propagation.}
The simplest way to call \acl{propagate-iso} to complete the isomorphism propagation is
with the same arguments as before, but just adding isomorphism information about
\acl{extend-path\-taken} and changing the \acl{:last-event}. This regenerates the events
from the previous call, but ACL2's redundancy checking makes this cheap. An alternative is
to also change the \acl{:first-event}, but then the tables from the previous call must be
given to the new call. There is an option to print these out in the right form for the
following call, but they can be large so it is much more concise and also less brittle to have
\acl{propagate-iso} regenerate them. 

For this drone planner,\ \acl{propagate-iso} generates 32 translated functions, 6
isomorphisms including defining 4 to-and-from translator functions, and 317 non-local
theorems, all of which are proven with the automatically generated hints.


\section{Related Work}

The idea of isomorphic data type transformations is not new.
In particular, the authors of this paper worked on
the design and implementation of a similar transformation
for the Specware system \cite{specware-www}.
The novel contributions of this paper are:
(i) the mathematical characterization of this transformation
with the `initiation' and `propagation' nomenclature
in \secref{sec:overview}; and
(ii) the realization in the ACL2 theorem prover
described in \secref{sec:tools}
and exemplified in \secref{sec:examples}.
Since the Specware language is typed and higher-order,
the Specware version of the transformation
differs from the APT/ACL2 version in some important aspects.
More broadly, the Specware transformation system,
which drew inspiration from the KIDS transformation system \cite{kids},
is the closest work to APT,%
\footnote{The generic `APT' refers to
the whole library of transformation tools,
not just the isomorphic type transformation ones.}
and in fact provided much inspiration to it.
A difference between Specware and APT is that
the latter is tightly integrated with a theorem prover (ACL2),
while the former had interfaces and translations to external theorem provers.

Automated type transformation tools exists
for typed higher-order theorem provers like Isabelle/HOL and Coq
\cite{isa-sepref,isa-containers,coq-fiat,coq-eal}.
These tools support richer mappings than isomorphisms,
but it is not immediately apparent
how they fit the paradigm in \secref{sec:overview}
of taking as inputs any%
\footnote{Provided that the applicability conditions hold, obviously.}
function $f$ and isomorphic mappings
and returning as outputs a function $f'$ and theorems relating $f$ and $f'$.
In particular, some of the referenced tools are built for infrastructures
consisting of refinement monads
and libraries of verified data types and relative operations.
The referenced tools and APT's isomorphic type transformation tools
may mutually benefit from incorporating some of each other's ideas,
but many technical aspects are bound to differ
due to the inherent differences between ACL2
and typed higher-order theorem provers.
More broadly, while the aforementioned refinement monads
follow a well-established refinement approach
based on inclusion of sets of behaviors,
APT is better suited to
the pop-refinement approach, mentioned in \secref{sec:intro},
based on inclusion of sets of implementations,
which in particular provides more flexibility and clarity
in matters of non-determinism, under-specification, and hyperproperties
\cite[Section 4.4]{popref}.

Smith \cite{smith-connections} developed
the concept of connections between theories,
which generates similar definitions for datatype functions,
but based on homomorphisms between types rather than isomorphisms.
Connections arose by generalizing from
transformations implemented in the KIDS system \cite{kids}.


\section{Future Work}
\label{sec:future}

\acl{Defiso} may already provide
all the features needed for isomorphic transformations.
Any future extensions may be driven by non-APT uses of this more general tool.

\acl{Isodata} should be extended to overcome the limitations noted in
\secref{sec:isodata}.  Eventually, \acl{isodata}
should be able to partition all the $n$ arguments and $m$ results of a function into
disjoint subsets, and apply a different isomorphic mapping to each subset (possibly the
identity one, for arguments and results to leave unchanged); each such isomorphic mapping
may map tuples (i.e.\ multiple arguments or results) to tuples of possibly different
sizes.  This will enable the use of \acl{isodata} for perhaps less expected
transformations, such as reordering arguments/results, renaming arguments, grouping multiple
arguments/results into single list arguments/results, ungrouping single list
arguments/results into multiple arguments/results, and adding redundant arguments
(caching) for incrementalizing computations.  The realization that so many apparently
different kinds of transformations can be unified under the \acl{isodata} umbrella,
contributes to the quest for a ``minimal'' and ``complete'' set of program
transformations; it is an open question whether such a set exists.

\acl{Propagate-iso} needs to be tested on more kinds of examples. There is likely
incremental improvement possible for the hints it produces. This is difficult in general
because most of the theorems produced need to be used as well as proved: adding hypotheses
can make the proof easier, but it makes them more difficult to use. Also, we would like to
generalize the automatic creation of dependent isomorphisms, which currently handle
predicates using list destructors. It should be
straightforward to additonally handle predicates using destructors such as those from
\acl{defaggregate}. Calling \acl{propagate-iso} could be made easier for the user if more
information were automatically extracted from theorems in the world.

While isomorphisms cover a wide variety of mappings,
many mappings of interest are not isomorphisms.
We have worked out a preliminary design for a transformation
that is similar to \acl{isodata} but allows richer mappings.
Some of its underlying concepts are similar to \acl{isodata},
but there are necessarily some new concepts as well.
In the aforementioned quest for a minimal and complete set of transformations,
\acl{isodata} may turn out to be a special case
of this upcoming new transformation.


\bibliographystyle{eptcs}
\bibliography{refs}


\end{document}